\documentclass[aps,prd,preprint,groupedaddress,showpacs,nofootinbib]{revtex4-1}

\usepackage{graphicx}
\usepackage{dcolumn}
\usepackage{bm}

\begin{document}


\title{Two Nucleon (B-L)-Conserving Reactions Involving Tau Leptons}
\author {Douglas Bryman}


\affiliation{Department of Physics and Astronomy, University of British Columbia, Vancouver, B.C. V6T 1Z1, Canada}


\date{\today}

\begin{abstract}
Tau lepton emission in  two-nucleon disappearance reactions from within nuclei 
which conserve baryon number minus lepton number
 (B-L) is considered. 
It is shown that some existing limits  on proton decay channels and two-nucleon 
disappearance reactions resulting in electrons and  muons can be applied to $\Delta B=\Delta L=2$
  decays involving $\tau$  leptons. For the two-nucleon disappearance channel $np\to\tau^+ \overline\nu_\tau$  
the  estimated limit for the partial mean  life is 
$\frac{\tau}{Br}>1\times10^{30}$ yrs based on results from the IMB3 experiment. Re-analysis of 
existing data and future experiments could result in higher sensitivity for two-nucleon disappearance 
modes involving $\tau$ lepton final states.

\end{abstract}


\maketitle

\section{Introduction}

The search for baryon number (B) violating reactions is motivated by
 the apparent dominance of matter over anti-matter in the universe.
 Although predictions for proton decay ($\Delta B=1$) from Grand
 Unified Theories have not been confirmed by experiments ruling out
 the minimal SU(5) scenario, baryon and lepton number violations occur
 in many extensions of the Standard Model (SM) at potentially
 observable rates\cite{Nath}.  
Theories incorporating new heavy particles and new interactions such as
supersymmetry or extra dimensions may also include baryon number violation.
In some extensions of the SM the
 quantity B-L (baryon number minus lepton number (L)) is conserved, so
 modes like $p\to e^+\pi^0$ with $\Delta B=\Delta L=1$ are
 favored.

(B-L)-conserving reactions with $\Delta B=\Delta L=2$ involving 
 two-nucleon disappearance from within nuclei such as   $pp\to e^+e^+$, $pp\to
\mu^+\mu^+$, $pp\to e^+\mu^+$, $np\to e^+ \overline\nu_e$ and $np\to
\mu^+ \overline\nu_{\mu}$ have been searched for \cite{Frej}.
The Particle Data Group (PDG)\cite{PDG} lists 13 two-nucleon
disappearance partial mean life 
limits 
 with the longest  $\frac{\tau}{Br}>5.8\times10^{30}$ 
yrs\cite{Frej}\footnote{All limits discussed here will be at 90\% c.l.} due to 
the channel $pp\to
e^+e^+$. The ``partial mean life'' limits are the limits on 
$\frac{\tau}{Br_i}$ where $\tau$ is the total mean life and $Br_i$ 
is the branching fraction for the decay mode in question\cite{PDG}.  
Processes with $\Delta B=\Delta L=2$ have been considered in theoretical models which typically
involved an extended Higgs sector.  For example, Arnellos and Marciano\cite{AM}
discussed such  processes in the context of an SU(5) model containing a Higgs
50-plet. Invisible modes with $\Delta B=\Delta L=2$, particularly $nn
\to \overline\nu_{l} \overline\nu_{l}$ where $l=e,\mu,\tau$, would be
severely suppressed by angular momentum conservation but two-nucleon
disappearance modes with $\Delta B=2$ and $\Delta L=0$ such as
$nn\to$invisible or $nn\to\nu_{l} \overline\nu_{l}$ have been searched
for \cite{Kam}; the resulting limits would also apply to
$nn\to\overline\nu_{l} \overline\nu_{l}$ decays. Recently, theoretical
models which do not generate proton decay but involve ~$\Delta B=2$
(which results in neutron-anti-neutron oscillations) and $\Delta
B=\Delta L=2$ have been considered by Arnold, Fornal, and Wise
\cite{AFW}. Dimensional analysis indicates that the fundamental scale for 
suppression of $\Delta B=2$ processes
may be at the few TeV scale\cite{MarshakMohapatra}, much lower than the 
grand-unification scale suggested by proton decay.

While single nucleon decay to final states involving on-shell $\tau$
leptons cannot occur due to energy conservation, $\Delta B=\Delta L=2$
two-nucleon disappearance reactions within nuclei such as 
 $np\to \tau^+ \overline\nu_{\tau}$ and  $pp\to
\tau^+ e^+$ could proceed. 
 Reactions with $\Delta B=\Delta L=2$ involving $\tau$ leptons may possibly be
enhanced over direct decays to final states with electrons and muons
due to generation-dependent effects or mass-dependent couplings to
exotic Higgs particles. For two-nucleon disappearance modes with
$\tau$  lepton final states, several $\tau$ decay channels may be 
observed  including $\tau^+ \to e^+ \nu_e \overline\nu_{\tau}$,
$\tau^+ \to \mu^+ \nu_{\mu} \overline\nu_{\tau}$, $\tau^+ \to \pi^+
\overline\nu_{\tau}$, and $\tau^+ \to \pi^+ \pi^0 \overline\nu_{\tau}$
with branching ratios 17.8\%, 17.4\%, 10.8\%, and 25.5\%,
respectively\cite{PDG}. B-violating decays with $\Delta B=1$ such as
$p \to \pi^+ \overline\nu_{\tau}$ involving $\tau$ leptons in virtual
intermediate states have also been discussed\cite{Nath} but the
effects were found to be very small.

At present, there are no partial mean life constraints on  two-nucleon decays into
$\tau$ leptons listed in the PDG compilation\cite{PDG}.  However, the
limits found by previous proton decay experiments may be useful for
obtaining first limits on such reactions. In particular, searches for 
direct decays to 
electrons, muons, or pions  used kinematic
constraints which may have  
overlap with
some $\tau$ lepton decay channels. For example, the results for $np\to e^+
\overline\nu_e$, $np\to \mu^+ \overline\nu_{\mu}$, $p \to e^+ \nu
\nu$, and $p \to \mu^+ \nu \nu$ may be used to obtain limits on
$np\to \tau^+ \overline\nu_{\tau}$ in which the $\tau$ decays to
leptons. Similarly, searches for $pp \to \mu^+ e^+$ and $pp \to e^+
e^+$ reactions may be examined for applicability to limits on $pp \to
\tau^+ e^+$ and the results for $pn \to\pi^+ \pi^0 $ searches may be
considered to limit the mode $np \to\tau^+ \overline\nu_{\tau}$ in
which the $\tau$ lepton decay $\tau^+ \to \pi^+ \pi^0
\overline\nu_{\tau}$ occurs. Limits obtained on single proton decay
$p\to\pi^+ \nu$ and $p\to \rho \nu$ may also be considered for
application to searches for $np \to \tau^+ \overline\nu_{\tau}$
followed by $\tau^+ \to \pi^+ \overline\nu_{\tau} $ and $\tau^+ \to
\pi^+ \pi^0 \overline\nu_{\tau}$.  In the following, the results from
several experiments will be examined to determine their applicability
for setting limits on the partial mean life  for the reaction
$np \to \tau^+ \overline\nu_{\tau} $.

\section{Experimental Limits on $np\to \tau^+ \overline\nu_{\tau}$}


The Frejus experiment\cite{Frej} with target material of Fe provided
limits on many two-nucleon disappearance modes.  
In this 
experiment, a minimum energy threshold of 200 MeV for triggering was
applied and the 
 energy resolution  was quoted as $\sigma_E=\frac{\Delta E^{em}}{E}=15 \%$
(12\%) at 400 (1000) MeV for electromagnetic showers due to electrons
or gamma rays\cite{Frej}.   For muons at 300 MeV/c traveling
perpendicular to the iron plates the momentum resolution was  3\% (or 10 MeV/c)\cite{Frej}. 
Accepted events were required to be consistent with kinematic
expectations such as  energy conservation. For example, considering the
disappearance of a neutron and proton from $^{56}Fe$, the mass  difference
between the ground states of $^{56}Fe$ and $^{54}Mn$\cite{NDC} results in 
1858 MeV available energy for the decay products. Using the $\tau$ lepton 
mass $m_{\tau}=1776.82\pm0.16$ MeV/$c^2$\cite{PDG}, the final state leptons in 
two-nucleon disappearance reactions
$np\to \tau^+ \overline\nu_{\tau}$  from $^{56}Fe$ would each have momentum approximately $P=79$
MeV/c. In the following, the effects of the small kinetic energy
of the $\tau$ lepton (T=1.8 MeV) will be neglected.

For $\tau$ lepton decays at rest to electrons and muons, $\tau^+ \to
e^+ \nu_e \overline\nu_{\tau}$ and $\tau^+ \to \mu^+ \nu_e
\overline\nu_{\tau}$, the momentum distribution of the charged lepton
decay products follows the ``Michel'' spectrum\cite{Michel}, originally
used to describe muon decays. For $\tau^+ \to
e^+ \nu_e \overline\nu_{\tau}$ ($\tau^+ \to \mu^+ \nu_e
\overline\nu_{\tau}$)  the maximum electron (muon)
momentum occurs  at approximately $P_{e(\mu)}=888$ $ (885)$ MeV/c. In order to find
limits on 
$np\to \tau^+ \overline\nu_{\tau}$ reactions using previous
 results obtained with  $e^+$ or $\mu^+$ decay products,
the fraction  of the $\tau^+ \to e^+ \nu_e \overline\nu_{\tau}$
and $\tau^+ \to \mu^+ \nu_e \overline\nu_{\tau}$ charged lepton 
decay distributions accepted by the kinematic regions used by the experiments 
needs to be
estimated. For example, in the
Frejus evaluation of  $n \to \pi^0
\nu$, an energy acceptance window of about $\pm 3\sigma_E$
relative to the expected  visible energy was employed. For the cases $np\to e^+
\overline\nu_e$ and $np\to \mu^+ \overline\nu_{\mu}$ comparable
visible energy acceptance windows would be approximately 595-1263 MeV 
 and 717-936 MeV, respectively.  For these energy windows approximately
$\epsilon_e=60\%$ of 
 $\tau^+\to e^+ \nu_e \overline\nu_{\tau}$ decays and $\epsilon_{\mu}=18\%$ 
     of $\tau^+ \to \mu^+ \nu_e
\overline\nu_{\tau}$ decays would have been accepted. 

Limits on the partial mean life of the decay $np \to \tau^+ \overline\nu_{\tau}$ can then be roughly
evaluated based on the Frejus limits on the reactions $np \to e^+
\overline\nu_{e}$, $\frac{\tau}{Br}>2.8\times10^{30}$ yrs, and $np \to \mu^+
\overline\nu_{\mu}$, $\frac{\tau}{Br}>1.6\times10^{30}$ yrs\cite{Frej}.
Using the $\tau^+ \to e^+ \nu_e
\overline\nu_{\tau}$ and $\tau^+ \to \mu^+ \nu_e \overline\nu_{\tau}$
branching fractions  and the  acceptance fractions $\epsilon_e$ and $\epsilon_{\mu}$
results in estimated
limits for $np \to \tau^+ \overline\nu_{\tau}$ reactions
$\frac{\tau}{Br}>0.3\times10^{30}$ yrs and 
$\frac{\tau}{Br}>0.05\times10^{30}$ yrs.  

The
Frejus experiment also produced the best limit on $p \to \mu^+ \nu
\nu$ decay, $\frac{\tau}{Br}>2.1\times 10^{31}$
yrs.  Here the maximum visible energy would be about 365 MeV. Using a
$\pm 3\sigma_E$ energy window results in an  acceptance fraction of $\epsilon_{\mu}=11\%$ 
for  $\tau^+ \to \mu^+ \nu_{\mu}
\overline\nu_{\tau}$ decays and an estimated lower limit on the
partial mean life for the reaction  $np \to \tau^+ \overline\nu_{\tau}$, 
$\frac{\tau}{Br}>0.4\times10^{30}$ yrs.

\begin{table*}[htb]%
\caption{Summary of results for the limits on the partial mean life $\frac{\tau}{Br}$ 
of two-nucleon disappearance
in the reaction $np \to \tau^+ \overline\nu_{\tau}$. The first column indicates
the original reaction searched for in the experiment referenced in the second column.
The third column shows the $\tau$ lepton decay channel used in the analysis.}
\begin{tabular} {@{}cccc}
\hline\hline
Decay channel    & Experiment    & $\tau$ decay    &   $\frac{\tau}{Br}$ [yrs] for $np \to \tau^+ \overline\nu_{\tau}$\\
\hline
$np\to e^+ \nu_e$ & Frejus\cite{Frej} & $\tau^+\to e^+ \nu_e \overline\nu_{\tau}$ & $>0.3\times10^{30}$ \\ [1ex]
\hline
$np\to \mu^+ \nu_{\mu}$ & Frejus\cite{Frej} & $\tau^+\to \mu^+ \nu_{\mu} \overline\nu_{\tau}$ & $>0.05\times10^{30}$ \\ [1ex]\hline
$p\to \mu^+ \nu \nu$ & Frejus\cite{Frej} & $\tau^+\to \mu^+ \nu_{\mu} \overline\nu_{\tau}$ & $>0.4\times10^{30}$ \\ [1ex]
\hline
$p\to e^+ \nu \nu$ & IMB3\cite{IMB3a} & $\tau^+\to e^+ \nu_e \overline\nu_{\tau}$ & $>1\times10^{30}$ \\ [1ex]

\hline
\end{tabular}%
\label{table:summary}
\end{table*}%

The IMB3 experiment\cite{IMB3a} gave the best limit on $p \to e^+ \nu
\nu$ decay, $\frac{\tau}{Br}>17\times10^{30}$ yrs,
which can be used to evaluate the process $np \to \tau^+
\overline\nu_{\tau}$ followed by $\tau \to e^+ \nu_e
\overline\nu_{\tau}$.  
IMB3 used energy conservation  and anisotropy selection criteria
 to identify  events compatible with the $p \to e^+ \nu
\nu$ decay signature. The anisotropy
of an event was a measure of  the net momentum. For $p \to e^+ \nu
\nu$ decay the anisotropy criterion  was chosen to accept
 single particle emission which would have been consistent 
with $\tau^+ \to e^+ \nu_e \overline\nu_{\tau}$ decay.
The energy region used to search for $p \to e^+
\nu \nu$\cite{IMB3b} was approximately 100-550 MeV resulting in an
acceptance fraction for $\tau^+ \to e^+ \nu_e \overline\nu_{\tau}$
decays of $\epsilon_e=32\%$.
The  estimated limit on the partial mean life for $np \to \tau^+ \overline\nu_{\tau}$ 
is then $\frac{\tau}{Br}>1\times10^{30}$ yrs.

Several other experimental results were considered for evaluating two-nucleon disappearance
reactions $np \to \tau^+ \overline\nu_{\tau}$ and $pp \to \tau^+ e^+$ using various $\tau$ decay
channels. These include the Soudan 2 limit on $p\to\pi^+ \nu$ decay\cite{Sou} using
$\tau \to \pi^+\overline\nu_{\tau}$, the IMB3
 limit on $p \to \rho^+ \nu$ decay\cite{IMB3a} using  $\tau^+ \to \rho^+
\overline\nu_{\tau}$, the Frejus limit on $pp\to e^+ e^+$
decay\cite{Frej} using $\tau^+ \to e^+ \nu_e \overline\nu_{\tau}$ and the  Super-Kamiokande
experiment result on 
$p\to\pi^+\overline\nu$\cite{SK-arxiv}.
However, it appears that
the selection criteria used to obtain these results, principally, the 
allowed energy regions, would have had little overlap with the kinematics of the 
decay products of the relevant 
$\tau$ decays. Limits on 
$nn \to \tau^{\pm} e^{\mp}$ were also  considered using results obtained
by IMB3\cite{IMB3b} on $n \to e^+ e^- \nu$ and  $n \to \mu^+ e^- \nu$; however, 
due to the low momentum of the  electron or positron 
from the reactions $nn \to \tau^{\pm} e^{\mp}$, the experimental anisotropy 
regions selected\cite{IMB3b} would   have had minimal overlap.

\section{Summary}

Table~\ref{table:summary} shows the 
lower limit estimates for the partial mean life of
the
$\Delta B=\Delta L=2$ process $np \to \tau^+ \overline\nu_{\tau}$ obtained 
by examining previous results of several nucleon decay experiments.
The largest lower limit  found on the partial mean life was 
$\frac{\tau}{Br}>1\times10^{30}$ yrs, 
based on the IMB3 experimental result for $p\to e^+ \nu
\nu$\cite{IMB3a}.
These results also apply to the (B+L)-conserving process $np \to \tau^+ \nu_{\tau}$.
 Re-analysis of data from Frejus, IMB3, Soudan 2, Super-Kamiokande, and
other nucleon decay experiments would likely result in considerably
greater sensitivity on two-nucleon disappearance reactions resulting in $\tau$ leptons.
Future and current nucleon decay experiments would
be enhanced by explicitly triggering on and considering such channels.
In particular, a major advance in sensitivity to two-nucleon disappearance channels
involving $\tau$ leptons could come from analysis of data from the Super-Kamiokande experiment
which might reach partial mean lives of $10^{33}$ yrs or more\cite{SK-arxiv}.

\section{Acknowledgments}
Thanks to T. Numao, L. Littenberg, and D. Morrissey for helpful comments.




\nocite{*}
\bibliographystyle{elsarticle-num}
\bibliography{martin}



\end{document}